\documentclass[conference]{IEEEtran}

\usepackage{cite}

\ifCLASSINFOpdf
   \usepackage[pdftex]{graphicx}
   \graphicspath{{figures/}}
\else
   \usepackage{graphicx}
\fi
\graphicspath{{figures/}}
\DeclareGraphicsExtensions{.eps,.pdf,.jpeg,.PNG}
\usepackage{epstopdf}

\usepackage{algorithm}
\usepackage{algcompatible}

\algnewcommand\algorithmicswitch{\textbf{switch}}
\algnewcommand\algorithmiccase{\textbf{case}}
\algnewcommand\algorithmicassert{\texttt{assert}}
\algnewcommand\Assert[1]{\State \algorithmicassert(#1)}%
\algdef{SE}[SWITCH]{Switch}{EndSwitch}[1]{\algorithmicswitch\ #1\ \algorithmicdo}   {\algorithmicend\ \algorithmicswitch}%
\algdef{SE}[CASE]{Case}{EndCase}[1]{\algorithmiccase\ #1}{\algorithmicend\     \algorithmiccase}%
\algtext*{EndSwitch}%
\algtext*{EndCase}%

\usepackage{amsmath}
\newcommand{\be}{\begin{equation}}
\newcommand{\ee}{\end{equation}}
\usepackage{array}
\usepackage{multirow}
\usepackage{booktabs}
\newcolumntype{L}[1]{>{\raggedright\let\newline\\\arraybackslash\hspace{0pt}}m{#1}}
\newcolumntype{C}[1]{>{\centering\let\newline\\\arraybackslash\hspace{0pt}}m{#1}}
\newcolumntype{R}[1]{>{\raggedleft\let\newline\\\arraybackslash\hspace{0pt}}m{#1}}

\usepackage[lofdepth,lotdepth]{subfig}

\usepackage[turkish]{babel}
\usepackage[utf8]{inputenc} 
\usepackage[T1]{fontenc}

\IEEEoverridecommandlockouts

\AtBeginDocument{%
	\renewcommand\tablename{TABLE}
}

\AtBeginDocument{%
	\renewcommand\figurename{Figure}
}

\AtBeginDocument{%
	\renewcommand\abstractname{Abstract}
}

\AtBeginDocument{%
	\renewcommand\refname{References}
}

\begin{document}
%
\title{\vspace*{-0.9cm}Instagram Fake and Automated Account Detection}
\author{\fontsize{13}{13}\IEEEauthorblockN{
		Fatih Cagatay Akyon \IEEEauthorrefmark{1},
		Esat Kalfaoglu \IEEEauthorrefmark{2}
	} \\
	\IEEEauthorblockA{\IEEEauthorrefmark{1}Electrical \& Electronics Engineering, Ihsan Dogramaci Bilkent University, Ankara, Turkey}
	\IEEEauthorblockA{\IEEEauthorrefmark{2}Electrical \& Electronics Engineering, Middle East Technical University, Ankara, Turkey}
	\{akyon\}@ee.bilkent.edu.tr, \{kalfaoglu\}@metu.edu.tr}

\IEEEpubid{\makebox[\columnwidth]{978-1-7281-2868-9/19/\$ 31.00 © 2019 IEEE\hfill}
\hspace{\columnsep}\makebox[\columnwidth]{}}

\maketitle

\begin{abstract}
	Fake engagement is one of the significant problems in Online Social Networks (OSNs) which is used to increase the popularity of an account in an inorganic manner. The detection of fake engagement is crucial because it leads to loss of money for businesses, wrong audience targeting in advertising, wrong product predictions systems, and unhealthy social network environment. This study is related with the detection of fake and automated accounts which leads to fake engagement on Instagram. Prior to this work, there were no publicly available dataset for fake and automated accounts. For this purpose, two datasets have been published for the detection of fake and automated accounts. For the detection of these accounts, machine learning algorithms like Naive Bayes, Logistic Regression, Support Vector Machines and Neural Networks are applied. Additionally, for the detection of automated accounts, cost sensitive genetic algorithm is proposed to handle the unnatural bias in the dataset. To deal with the unevenness problem in the fake dataset, Smote-nc algorithm is implemented. For the automated and fake account detection datasets, 86\% and 96\% classification accuracies are obtained, respectively.
	
\end{abstract}

\begin{ozet}
	Sahte etkileşim, bir hesabın popülaritesini artırmak için çevrimiçi sosyal ağlarda kullanılan önemli sorunlardan bir tanesidir. Sahte etkileşim tespiti çok önemlidir, çünkü işletmeler için para kaybına, reklamlarda yanlış hedef kitleye yönelmeye, yanlış ürün tahmin sistemlerine ve sağlıksız sosyal ağ ortamına neden olur. Bu çalışma Instagram'da sahte etkileşime yol açan sahte ve otomatik hesapların tespiti ile ilgilidir. Bildiğimiz kadarıyla, sahte ve otomatik hesaplar için literatürde yayınlanmış bir veri seti bulunmamaktadır. Bu amaçla, sahte ve otomatik hesapların tespiti için iki veri seti oluşturulmuştur. Bu hesapların tespiti için Naive Bayes, Lojistik Regresyon, Destek Vektör Makineleri ve Sinir Ağları gibi makine öğrenme algoritmaları kullanılmıştır. Ek olarak, otomatik hesapların tespiti için, veri setindeki doğal olmayan sapmalar nedeniyle ceza puanlı genetik algoritma uygulanmıştır. Sahte hesap veri setindeki eşitsiz dağılım problemiyle başa çıkmak için ise SMOTE-NC algoritması uygulanmıştır. Otomatik ve sahte hesap algılama tespitinde sırasıyla \% 86 ve \%96 oranlarında başarım elde edilmiştir.

\end{ozet}
\begin{IEEEanahtar}
Sahte Etkileşim, makine öğrenmesi, çevrimiçi sosyal ağları, Instagram, genetik algoritma, Smote
\end{IEEEanahtar}
\IEEEpeerreviewmaketitle


%
\IEEEpeerreviewmaketitle

\section{Introduction}

Online Social Networks(OSNs) like Facebook and Instagram have becoming more and more popular and become the crucial part in Today's World. Beside the usage of OSNs as a medium of communication, they are also used to gain popularity and promote businesses. At the first glance, the popularity of an account is measured by some metrics like follower count or the properties of the shared contents like the number of likes, comments or views. Therefore, users of any social platforms might have a tendency to bolster its metrics in an artificial manner to get more benefits from OSNs.   

There are some common ways to increase the reputation of an account in social media. These ways can be listed as usage of bots, buying social metrics such as like, comment and follower, and usage of some platforms or networks which enables users to trade metrics \cite{sen2018worth}. A bot is a piece of software that completes automated tasks over the Internet. By a 2018 study done by Ghost Data, nearly 95 million Instagram accounts are automated \cite{growingbotproblem2018}. In 2016, bots generated more Internet traffic than humans \cite{efthimion2018supervised}. Additionally, by the creation of fake accounts, vendors sells likes and followers very easily. For example, a company called IDigic sells 50k followers for only 250 dollars \cite{fraudReport}.

All of these actions listed above are inorganic and termed as fake engagement. In other words, fake engagement term covers all types of automated activities such as liking and commenting on posts, following accounts, uploading posts/stories. In addition, buying social media metrics can also be included in fake engagement terminology. The detection of users who inorganically grow its account is significant because inorganic growth makes businesses pay more to users than its worth for advertising, makes advertisers reach to wrong audiences, make recommendation systems work inefficiently, make access to quality services and product harder. 

Fake engagement are divided into 2 separate topics which are the detection of automated accounts or bot accounts and fake accounts. As explained before, bot accounts are the users who performs automated activities like following users and liking media from related audience to increase its popularity metrics. Fake accounts are the accounts which are used to boost the social media metrics of a specific account who pays for this service. To highlight it more clearly, it can be also mentioned as fake followers. The main difference of automated and fake accounts is that automated account improves the metrics of itself while fake accounts improves the metrics of other users and creates unhealthy social media environment.


In the literature, there are some works and released datasets about the detection of fake engagement activity itself and the detection of users who engages inorganic activity in OSNs like Facebook and Twitter. The detection of Twitter fake accounts are studied in \cite{efthimion2018supervised} using support vector machines and logistic regression, in\cite{mohammadrezaei2018identifying} using graph based methods, in \cite{ercsahin2017twitter} using the joint usage of naive Bayes classifier with entropy minimization discretization. In \cite{el2016fake}, fake account detection on Twitter is studied by applying the GAIN measure \cite{Karegowda2010COMPARATIVESO} for weighting the all features used in the literature for this task and the improvements of such weighting on machine learning algorithms are shown. \cite{raturi2018machine} also focuses on the detection of fake accounts using NLP and machine learning tools but also proposes some security architectures and focuces also on Facebook. In \cite{cresci2015fame}, the main focus is on the detection of fake followers on Twitter with some machine learning algorithms. In \cite{li2016world}, fake social engagement on Youtube is studied by a graph diffusion process via local spectral subspace.

There are also some works in which Instagram are studied from the fake engagement perspective. Instagram has become one of the top social platforms. Instagram has reached about 1 billion monthly active user and 2 million monthly advertisers and users like 4.2 billions posts daily \cite{todd_2019}. Therefore, it is crucial to preserve the healthy environment in such an important social platform. In \cite{sen2018worth}, fake likes are tried to be determined on Instagram. In this study, the main concern is to estimate what is the probability that a user can like the post of another user based on the network closeness, interest overlap, liking frequency, influencer effect and link farming hashtag effect. \cite{spam} and \cite{comment} are about the detection of spammy posts and spammy comments on Instagram, respectively. In \cite{malicious}, Facebook employees studied the detection of malicious accounts from the requests sent on Facebook and Instagram but these method is not applicable with publicly available information because requests are reachable only from Facebook. From all of these studies, it is observed that there is not any work done regarding fake and automated account detection for Instagram with publicly available information; moreover, no publicly available dataset is present required for these analysis. In this work, we collect and annotate fake account and automated account datasets and present a detailed analysis on fake and automated account detection for Instagram using machine learning algorithms and explain the steps required for preprocessing. The dataset is available on https://github.com/fcakyon/instafake-dataset.

For the rest of the paper, in Section II, fake account detection dataset and features are detailed. In Section III, automated account detection dataset and features and cost sensitive feature selection algorithm is given. In Section IV, implemented classification algorithms are detailed. Section V presents the results and section VI concludes the paper.



\section{Fake Account Detection}
This section is related with the detection of fake accounts. Fake accounts are the accounts which are used to increase the popularity metrics of other users. For this reason, they have a tendency to have a high following and low follower counts. Their liking behavior may look randomly. The absence of profile picture and strange user names are the common characteristics of fake accounts. 

In this section, the dataset and the selected features for the dataset have been introduced for the detection of fake account. Then, the oversampling method is explained which is necessary for the unevenness in the number of real and fake account in the dataset.

\begin{figure}[t]
	\centering
	\shorthandoff{=}
	{\includegraphics[width=2.7in]{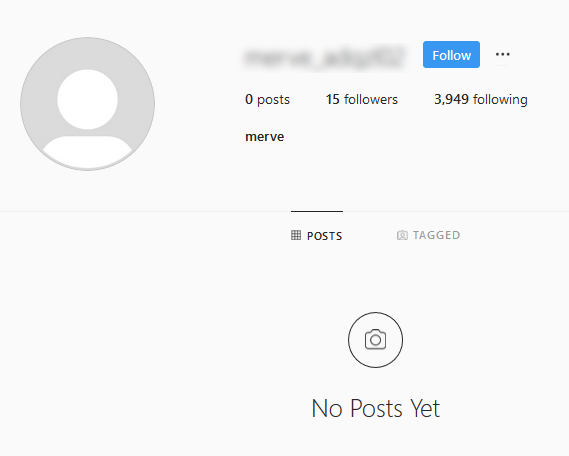}}
	\shorthandon{=}
	\caption{Fake account example from dataset. Most suspicious accounts are labeled as fake by hand.} 
	\label{fig:fkeaccount}
\end{figure}

\subsection{Dataset and Features}
For the dataset, 1002 real account and 201 fake account has been gathered after extensive manual labeling, including accounts from different countries and fields. During this gathering process, the points that is paid attention are follower and following counts, media counts, media posting dates or frequency, comments on media, some of the followed and following account, the existence of profile picture and the username of the profile. 

An example fake profile from the dataset can be seen in Figure \ref{fig:fkeaccount}. As seen, it has a high following number of 3949, and low follower number of 15, has no profile picture and no posted media. 

In the dataset, the selected base features can be listed as below:

\begin{itemize}
	\item{Total media number of the account.} 
	\item{Follower count of the account.}
	\item{Following count of the account.}
	\item{Number of digits present in account username.}
	\item{Whether account is private, or not (binary feature).}
\end{itemize}

To emphasize, all the features are not related with the user media, therefore the algorithm is not affected by the account privacy. The reason to add number of digits present in account username is that during the generation of fake accounts, some accounts are produced by adding different numbers to the same name. The number of digits distribution can be seen in Table \ref{Table:1}. As seen, more that 50\% of fake accounts have more than one digits while real account has no digits with about 89\%.

\begin{figure}[t]
	\centering
	\shorthandoff{=}
	{\includegraphics[width=3.0in]{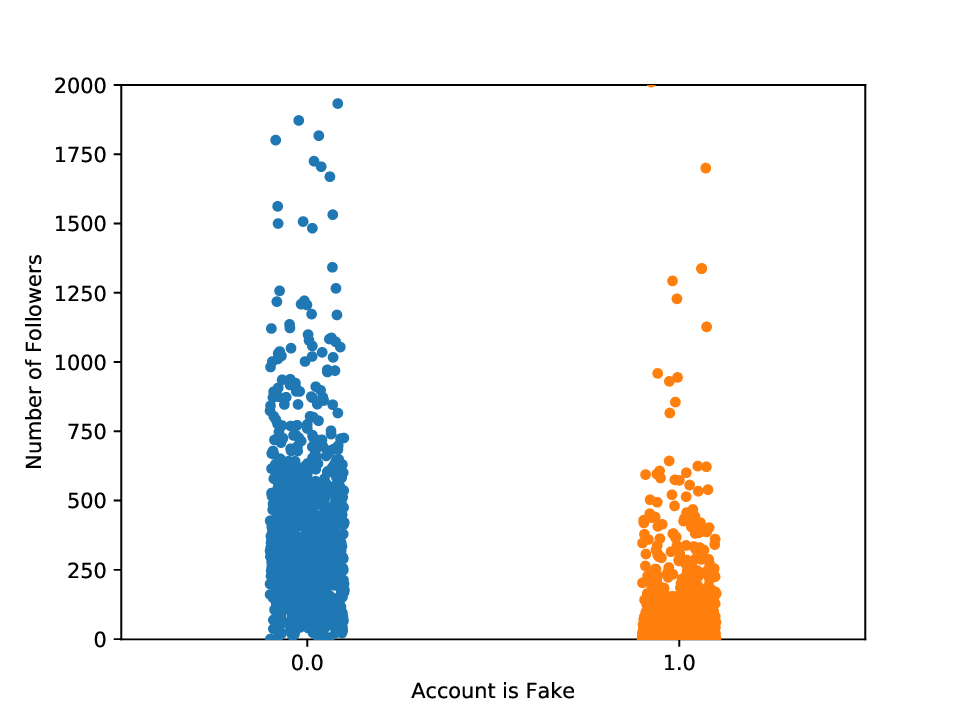}}
	\shorthandon{=}
	\caption{In-class data distributions for "follower count" feature.} 
	\label{fig:follower_fakeaccount}
\end{figure}

\begin{figure}[t]
	\centering
	\shorthandoff{=}
	{\includegraphics[width=3.0in]{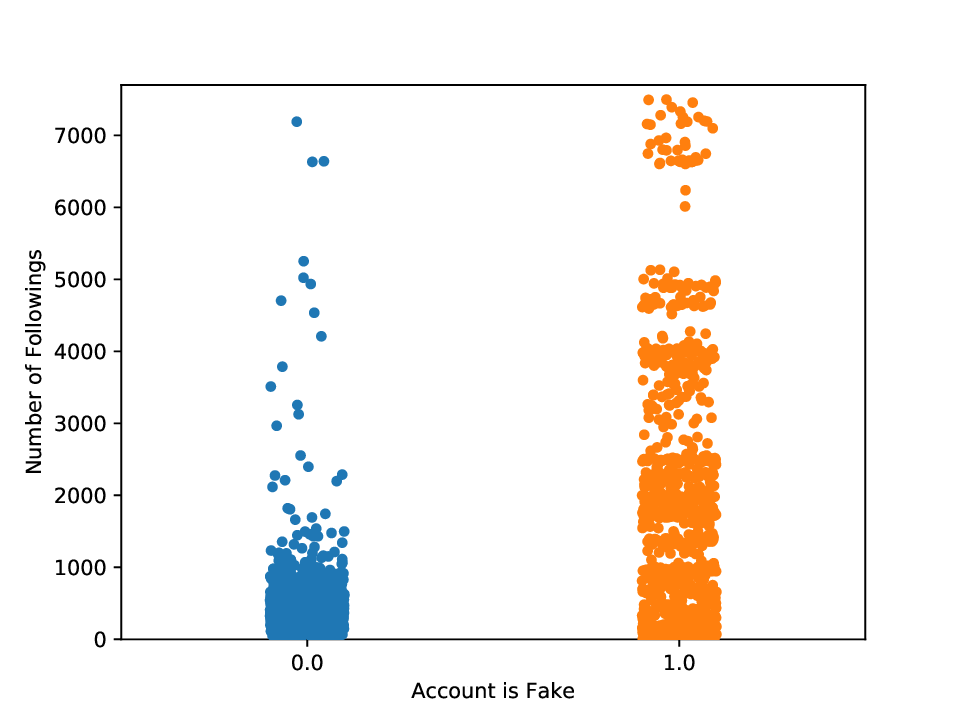}}
	\shorthandon{=}
	\caption{In-class data distributions for "following count" feature.} 
	\label{fig:following_fakeaccount}
\end{figure}

\begin{table}[t]
	\centering
	\caption{Distribution of accounts with changing number of digits included in their usernames.}
	\label{Table:number_of_digits}
	\renewcommand{\arraystretch}{1.2}
	\begin{tabular}[t]{C{2.5cm}C{1.5cm}C{1.5cm}}
		\toprule
		\textbf{\# of digits}&\textbf{Real accounts}&\textbf{Fake account}\\
		\midrule
		\textbf{0}	&88.9\% & 46.8\%\\
		\textbf{1}	&2.5\% & 10.0\%\\
		\textbf{2}	&5.3\% & 13.9\%\\
		\textbf{3}	&0.7\% & 11.4\%\\
		\textbf{3+}	&2.6\% & 17.9\%\\
		\bottomrule
	\end{tabular}
\end{table} 

\begin{figure}[t]
	\centering
	\shorthandoff{=}
	{\includegraphics[width=2.7in]{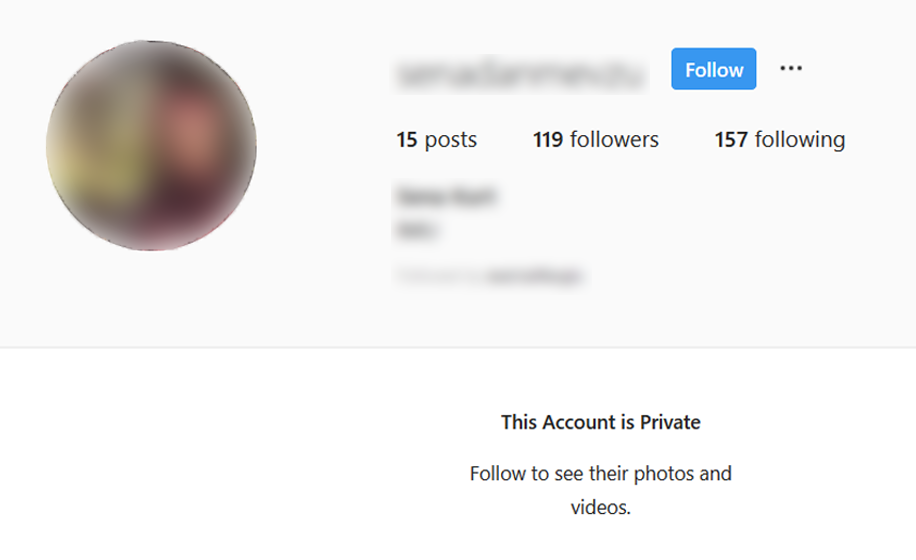}}
	\shorthandon{=}
	\caption{Example private account preview. Media details are not visible for private accounts.} 
	\label{fig:private_account}
\end{figure}

\subsection{Oversampling}
Distribution of classes in the fake account dataset is not even. This results in poor performance for the outnumbered class. SMOTE oversampling technique \cite{chawla2002smote} is utilized to increase number of samples for fake accounts. K is chosen as 5 for this work. In the implementation of SMOTE, SMOTE-NC is applied which considers not only the quantity classes but also the the categorical classes. After applying oversampling, all classifiers are trained on equal number of training samples per class (1002 per class).

\section{Automated Account Detection}
This section is related with the detection of automated accounts. Automated accounts which are also known as bot are the accounts which performs automated activities such as following, liking and commenting by targeting specific hashtags, locations of followers of specific accounts to increase their popularity metrics. Automated accounts might show fully inorganic behavior or organic and inorganic behavior together. The reason to observe organic behavior from such an account derives from the fact that user may continue to follow its own interests while the bot is running in background. 

In this section, three subsection are presented which are Dataset and Features, Bias Problem and the Cost Sensitive Feature Selection. Because of the bias problem in the generated dataset, generic algorithm is applied with the weighting of selected features which is detailed in Cost Sensitive Feature Selection.

\subsection{Dataset and Features}
\begin{figure}[t]
	\centering
	\shorthandoff{=}
	{\includegraphics[width=3.0in]{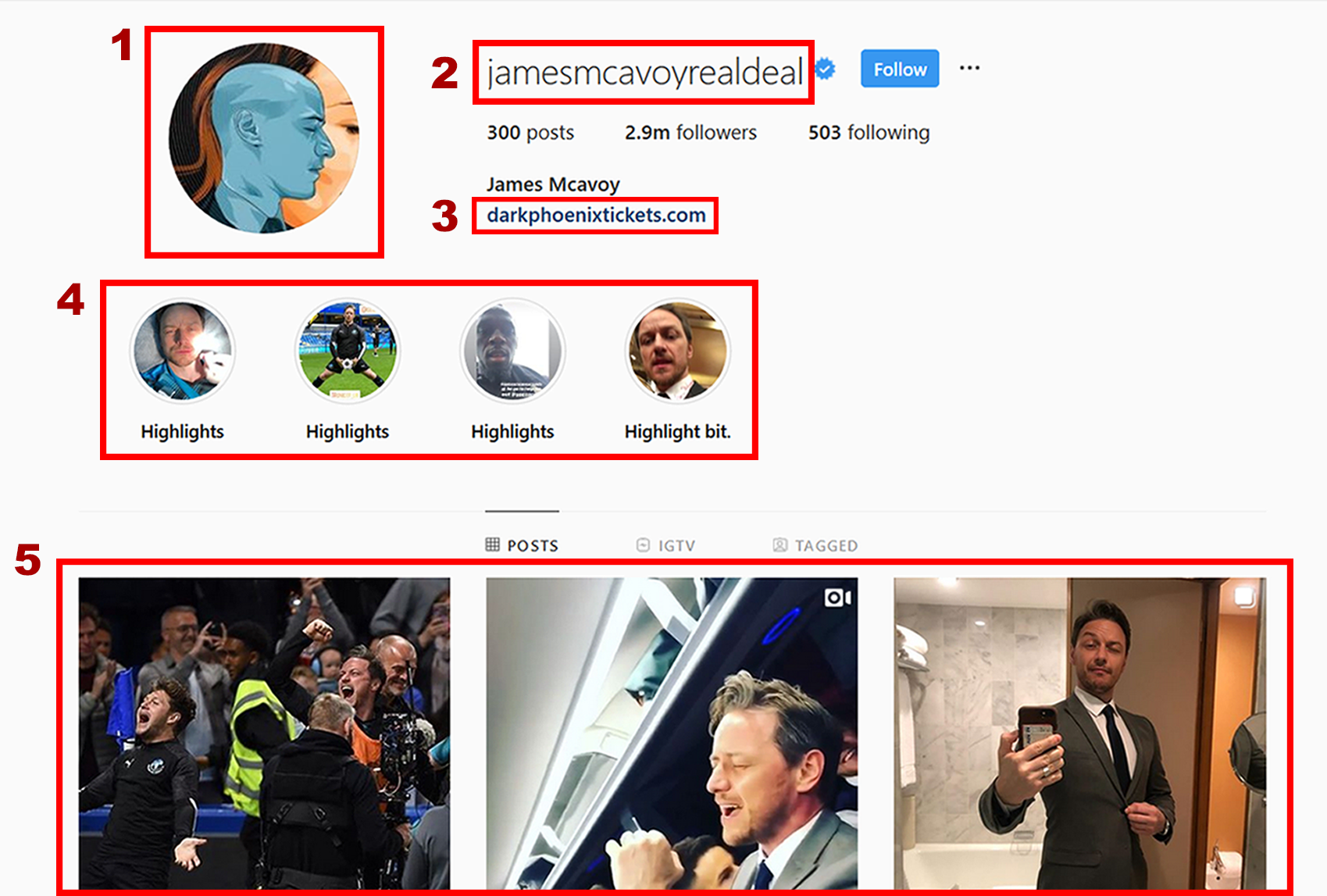}}
	\shorthandon{=}
	\caption{General preview of an Instagram profile. 1: Profile picture, 2: Username, 3: External URL, 4: Highlight reel, 5: User media} 
	\label{fig:profile_preview}
\end{figure}
The dataset consists of 700 real account and 700 automated account gathered from different countries and fields. For the collection of real account, we have selected the people we know from our circle of friends. To collect the automated accounts, we examined the source codes of the most popular open source Instagram bots that form the essential portion of fake engagements and noted specific behaviors to tag these accounts. Hashtags are one of the most common ways for inorganic activity. For example, one of the criteria to detect fake likes on Instagram is to investigate like and follow trading hashtag usage \cite{sen2018worth}. In our case, most popular hashtags are targeted because it is observed that catching automated behaviour is more easy and faster. From these hastags, if a user follows and unfollows after predefined exact durations (as observed from online instagram automations tool parameter sets), it is labelled as automated account. Instagram API is used with a Python wrapper to collect detailed media and user information of the accounts over 6 months period. For privacy reasons, any user related info is discarded for this work (user names, user photos, comment contents, hashtag names etc.).


Below are the scrapped base features from these accounts:
\begin{itemize}
	\item{Total media number of the accounts.} 
	\item{Follower count of the account.}
	\item{Following count of the account.}
	\item{Whether account has at least one highlight reel, or not (binary feature).}
	\item{Whether account has external url in porfile, or not (binary feature).}
	\item{Number of photos user is tagged by someone else.}
	\item{Average recent media hashtag number.}
\end{itemize}
If the account has no media, all features scrapped from user posts are assigned as 0. Furthermore, additional helpful features are derived using the base features such as:
\begin{itemize}
	\item{Average recent media like to comment ratio (LCR)}
	\item{Follower to following ratio (FFR).}
	\item{Whether account has not any media, or not (binary feature).}
\end{itemize}

Here, "recent" means, the corresponding features are scrapped/calculated using only the media information that is posted in the last 18 months. To understand some of the features like highlight reel or private profile, Figure \ref{fig:private_account}  and Figure \ref{fig:profile_preview} can be examined. To emphasize, the proposed automated account detection necessitates the access to user media. 

\subsection{Bias Problem}
\begin{figure}[t]
	\centering
	\shorthandoff{=}
	{\includegraphics[width=3.0in]{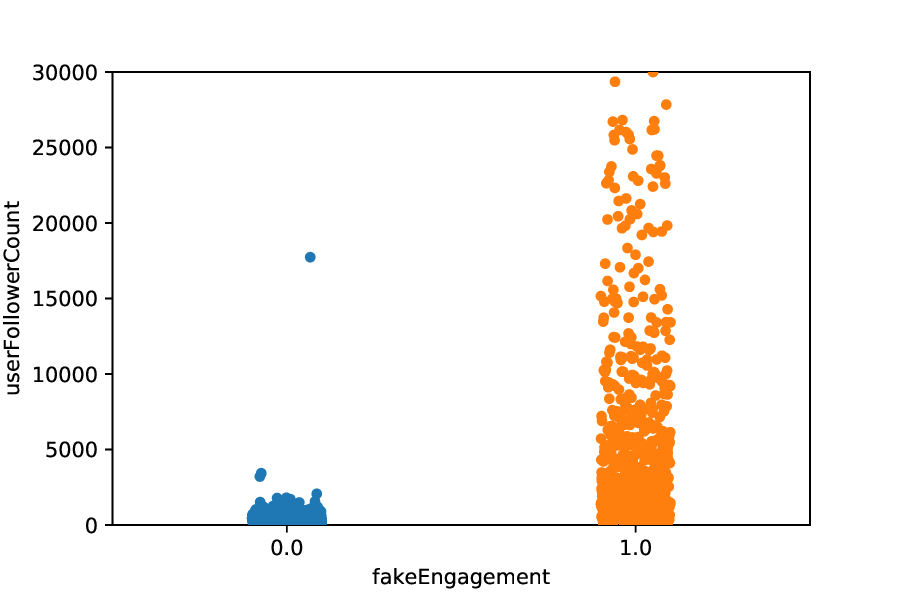}}
	\shorthandon{=}
	\caption{In-class data distributions for "follower count" feature.} 
	\label{fig:2}
\end{figure}

\begin{figure}[t]
	\centering
	\shorthandoff{=}
	{\includegraphics[width=3.0in]{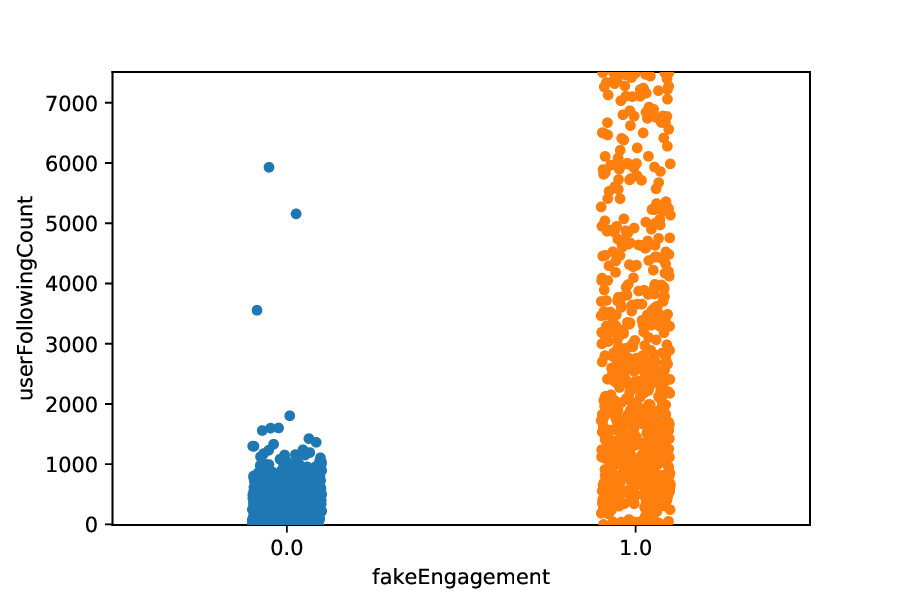}}
	\shorthandon{=}
	\caption{In-class data distributions for "following count" feature.} 
	\label{fig:3}
\end{figure}

\begin{figure}[!htb]
	\centering
	\shorthandoff{=}
	{\includegraphics[width=3.0in]{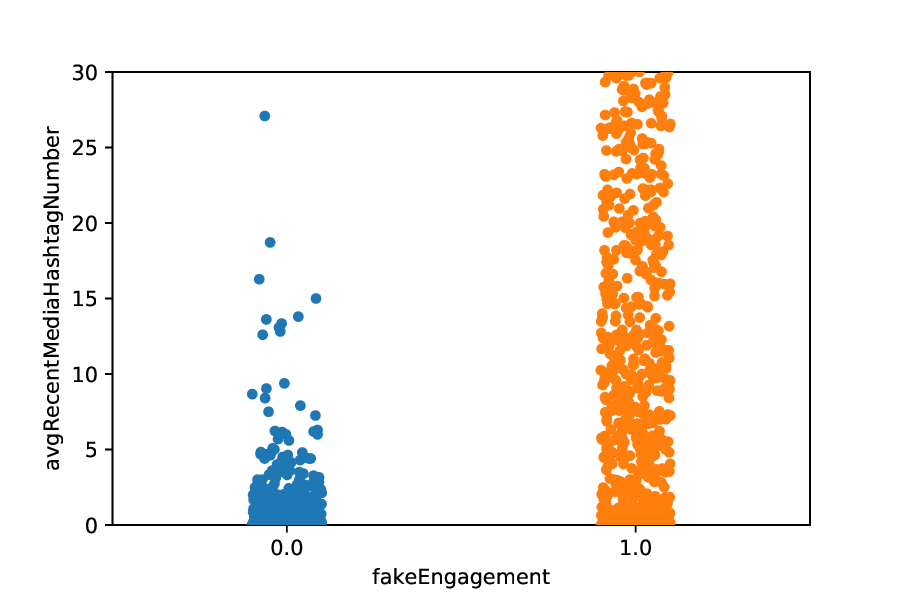}}
	\shorthandon{=}
	\caption{In-class data distributions for "average hashtag number" feature.} 
	\label{fig:4}
\end{figure}

There was some negative (unrealistic) bias present in some of the features. In Figures \ref{fig:2}-\ref{fig:4}, in-class distribution of the whole dataset with respect to chosen continuous features are illustrated. "Fake Engagement" being 1 correspond to the accounts involved in fake engagement or automated account while 0 correspond to the accounts that only has natural engagements or real account. As seen from the figures, chosen features have bias over the dataset. Although the bias in follower and following numbers is unrealistic (we have undeliberately chosen accounts with low follower\&following numbers as real accounts, it does not reflect the real situation), bias in average hashtag number is nearly natural (accounts with automated behaviour tend to use more hashtags per post). 

In Tables \ref{Table:1}-\ref{Table:2}, projection of the dataset over chosen binary features are given. As can be seen from tables, there are also bias present over these features, however this time; these are realistic bias. Highlight reels can be considered as an effective separator while real engagement accounts mostly not having URL present can also considered to be a true bias.


\begin{table}[t]
	\centering
	\caption{In-class data distributions for "user has highligh reels" feature.}
	\label{Table:1}
	\renewcommand{\arraystretch}{1.2}
	\begin{tabular}[t]{C{2.5cm}C{1.5cm}C{1.5cm}}
		\toprule
		&\textbf{Non-Automated Account}&\textbf{Automated Account}\\
		\midrule
		\textbf{Has Not Any Highligh Reels}	&468 & 260\\
		\midrule
		\textbf{Has Highligh Reels}	&232 &440\\
		\bottomrule
	\end{tabular}
\end{table}

\begin{table}[t]
	\centering
	\caption{In-class data distribution for "user has external URL" feature.}
	\label{Table:2}
	\renewcommand{\arraystretch}{1.2}
	\begin{tabular}[t]{C{2.5cm}C{1.5cm}C{1.5cm}}
		\toprule
		&\textbf{Non-Automated Account}&\textbf{Automated Account}\\
		\midrule
		\textbf{Has Not Any External URL}	&654 & 317\\
		\midrule
		\textbf{Has External URL}	&46 &383\\
		\bottomrule
	\end{tabular}
\end{table} 

\subsection{Cost Sensitive Feature Selection}
To overcome these unrealistic biases and select the most effective features, a cost sensitive genetic feature selection algorithm is developed. Pseudo code for the this genetic algorithm can be seen in Algorithm \ref{alg:ga}. Firstly, normalization is applied for the continuous features while binary features remained the same. Then, the normalized data is given to the genetic algorithm for a cost sensitive feature selection. An individual, whose length is same as total feature number of the data, is an array consisting of 1's and 0's (depending on if the feature is selected or not). To illustrate, if the second element of the individual is 1, then it means the second feature is one of the selected features for this specific individual. Using this representation, randomly generated individuals are used to form a population.

\begin{equation}
\label{eq:fitness}
Fitness = F2\ Score - 2 \times Tot. Feat. Cost
\end{equation}
\begin{equation}
\label{eq:f2}
F2\ Score= 5 \times \frac{precision \times recall}{4 \times precision + recall }
\end{equation}

Fitness calculation formula is given in Eq. \ref{eq:fitness}. Here $Tot. Feat. Cost$ is calculated by summing the individual costs of the selected features. Feature costs can be seen in Table \ref{Table:feat_cost}. These costs are determined based on the reliability of the data collection which is discussed in the previous Bias Problem section. Realistic biases are represented with lower features costs while the negative bias is represented with higher costs. For the F2 score in Eq. \ref{eq:fitness}, it belongs the classifier loss. F2 score formula is given in Eq. \ref{eq:f2}. For the calculation of F2 score, a two layer neural network architecture has been implemented which is detailed in Table \ref{Table:nn_details}.  Then, the selected features are used as an input to the all classification methods which will be detailed in the Classification Methods section. 

\begin{table}[t]
	\centering
	\caption{Costs of the features based on their bias reliability.}
	\label{Table:feat_cost}
	\renewcommand{\arraystretch}{1.2}
	\begin{tabular}[t]{C{3.0cm}C{2.0cm}}
		\toprule
		\textbf{Features}&\textbf{Cost}\\
		\midrule
		\text{Total media number}&$2$\\
		\text{Follower count}&$4$\\
		\text{Following count}&$4$\\
		\text{Has highlight reel}&$2$\\
		\text{Has external url}&$2$\\
		\text{Tag number}&$3$\\
		\text{Average hashtag number}&$2$\\
		\text{Has 0 media}&$1$\\
		\text{LCR}&$2$\\
		\text{FFR}&$4$\\
		\bottomrule
	\end{tabular}
	\vspace{5pt}
\end{table}

As the genetic operations; elitism, randomness, tournament based crossover, and mutation operations are implemented. At each generation, the individual having best fitness is directly selected for the next generation. 

\begin{table}[t]
	\centering
	\caption{Neural Network Details.}
	\label{Table:nn_details}
	\renewcommand{\arraystretch}{1.2}
	\begin{tabular}[t]{R{4.0cm}L{3.5cm}}
		\toprule
		\textbf{\# of Layers:} 			&$2$\\
		\textbf{\# of Hidden Units (per layer):} 	&$32$\\
		\textbf{Optimization:} 			&$ADAM\ with\ Minibatch$\\
		\textbf{Non-linearity:} 		&$ReLu$\\
		\textbf{Loss Function:} 		&$Categorical\ Crossentropy$\\
		\textbf{Learning Rate:} 		&$0.001$\\
		\textbf{Minibatch Size:} 		&$64$\\
		\textbf{Epochs:} 			&$100$\\
		\textbf{Train-Test Split:} 			&$\%70-\%30$\\
		\bottomrule
	\end{tabular}
\end{table}


\begin{algorithm}[t]
	\floatname{algorithm}{Algorithm}
	\caption{Genetic Algorithm}\label{alg:ga}
	{\fontsize{9}{9}\selectfont
		\begin{algorithmic}[1]
			\REQUIRE{$\mathbf{FullDataset}$: \textit{(Normalized) Dataset containing all of the features}, $\mathbf{Population Size}$: \textit{Number of individuals present at each generation}, $\mathbf{Number of Generations}$: \textit{Number of generations to be iterated on}, $\mathbf{Mutation Rate}$: \textit{Mutation probability}}
			\ENSURE{$\mathbf{ReducedDataset}$: \textit{Dataset containing only the selected features}}
			\STATEx{$\mathbf{Initializations:} \text{Initalize} \ \mathbf{Population} \ \text{randomly}$}
			\FOR{$ind = 1: \mathbf{Number of Generations}$}
			\STATE{$\text{Calculate}\ \mathbf{Population} \ \text{fitness}$}
			\STATE{$\text{Select the best individual (elitisim)}$}
			\STATE{$\text{Select 1 random individual (randomness)}$}
			\STATE{$\text{Perform xover to rest of the individuals (tournament)}$}
			\STATE{$\text{Mutate 1 individual with prob.}\ \mathbf{Mutation Rate}$}
			\STATE{$\text{Update}\ \mathbf{Population}$}
			\ENDFOR
			\STATE{$\text{Form}\ \mathbf{ReducedDataset}\ \text{using}\ \mathbf{Population}$}
			\STATEx{$\mathbf{return}\,\mathbf{ReducedDataset}$}
	\end{algorithmic}}
	
\end{algorithm}	

\begin{figure}[t]
	\centering
	\shorthandoff{=}
	{\includegraphics[width=3.0in]{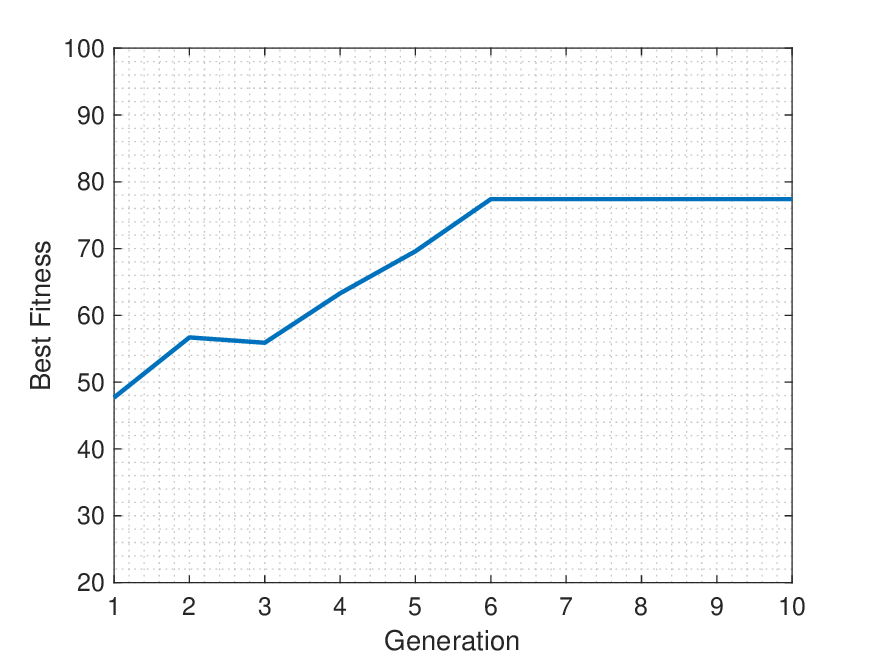}}
	\shorthandon{=}
	\caption{Fitness of the fittest individual of each generation.} 
	\label{fig:fitness_vs_gen}
\end{figure}

Calculated fitness values of the fittest individual of the population for a given generation can be seen in Figure \ref{fig:fitness_vs_gen}. As expected, evolution results in monotonic increase in the best fitness (considering mutation rate is very low). After 10 generations, individual with the best fitness value is used to select the best features. The selected features and total cost for selecting these features are given in Table \ref{Table:feat_cost}. 

\begin{table}[t]
	\centering
	\caption{Corresponding features of the fittest individual and their cost.}
	\label{Table:final_feat_cost}
	\renewcommand{\arraystretch}{1.2}
	\begin{tabular}[t]{C{2.5cm}C{2.0cm}}
		\toprule
		\textbf{Selected Features}&\textbf{Cost}\\
		\midrule
		\text{Total media number}&$2$\\
		\text{Has external url}&$2$\\
		\text{Average hashtag number}&$2$\\
		\text{LCR}&$1$\\
		\textbf{Total}&$\textbf{7}$\\
		\bottomrule
	\end{tabular}
	\vspace{5pt}
\end{table}

\section{Classification Methods}
Several traditional and neural network based learning methods are implemented as classifiers. As traditional methods, Naive Bayes, logistic regression and support vector machine (SVM) is
employed. In Naïve Bayes method, independent features of different classes are exploited to form the
posterior distributions of the classes, and maximum a-posteriori (MAP) estimation is performed. Logistic regression again
exploits independent features to differentiate two classes. SVM focuses on finding a hyperplane which separates a dataset in the best way. In addition to preprocessed data(features), raw
data can also be used as inputs while training and testing these type of networks.

\section{Results}
Through utilization of different kinds of algorithms, it is aimed to exploit different aspects of dataset (i.e., independence, separability, complex relations) which has not been deeply considered in literature and to find a good way of detection of the fake and automated accounts of Instagram. 

For the detection of automated accounts, cost sensitive selected features given in Table \ref{Table:final_feat_cost} have been used. For the detection of fake accounts, the base features of the fake-real dataset has been used directly.  

For the detection of automated accounts, to compare and test the effectiveness of the implemented techniques; Precision, Recall and  F1 Score are used as the evaluation metric as given in equations \ref{eq:Precision}-\ref{eq:F1 Score} respectively. Terms TP, TN, FP, and FN present in these equations correspond to True Positive, True Negative, False Positive and False Negative. F1 Score is more meaningful for performance evaluation because precision ignores the effect of FN and recall ignores the effect of FP. F1 score considers both of them. 

For a fair comparison, parameter optimization is performed by grid search for the classifiers that rely on parameters. Extensive 10-fold cross validations performed over the training portion of the dataset. Kernel is chosen as $radial\ basis\ function$, the $gamma$ parameter (kernel coefficient) is chosen as $1$, and the penalty parameter $C$ is chosen as $100$ (mid level regularization) for the SVM. $Solver$ is chosen as $Newton\ Conjugate\ Gradient$, $inverse\ of\ regularization\ coefficient$ (smaller value corresponds to stronger regularization) is chosen as $1000$ (tighter regularization), and $convergence\ tolerance\ level$ is chosen as $0.1$ for the logistic regression technique. Neural network is run with the parameters given in Table \ref{Table:nn_details}.

Test results for fake account detection dataset can be found in Table \ref{Table:result_fakeaccount_oversampling}. As mentioned before, SMOTE-NC has been used for the oversampling. F1 scores are calculated by the macro average method. The reason to use macro average is the fact that the distribution in the data does not reflect the real distribution in the Instagram.  It is desired to place importance equally on fake and real users. From the table, it is observed that oversampling has increased the performance of all methods. The highest performance without oversampling is observed with the neural network, while SVM and neural network perform equally in oversampling case with 94\%. 

Test results for automated account detection dataset can be found in Table \ref{Table:result_automated}. Neural network and SVM has the best overall F-1 scores. It is expected since it is known that neural networks can learn complex mappings with enough training data and SVMs are well known for optimizing the margin better than most other algorithms in binary tasks. Poor performance from Naive Bayes and logistic regression was not surprising since the features are not distinctly independent (considering these methods highly rely on the independence of the features). Moreover, low precision present for the Naive Bayes with Gaussian distribution means that the true distribution of the labels does not corporate with this likelihood assumption. 


\begin{equation}
\label{eq:Precision}
\text{Precision} = \frac{TP}{TP+FP}
\end{equation}

\begin{equation}
\label{eq:Recall}
\text{Recall} = \frac{TP}{TP+FN}
\end{equation}

\begin{equation}
\label{eq:F1 Score}
\text{F1 Score} = 2\times\frac{Recall \times Precision}{Recall + Precision}
\end{equation}

\begin{table}[th]
\centering
\caption{Evaluation of the classifiers over the fake account dataset.}
\label{Table:result_fakeaccount_oversampling}
\renewcommand{\arraystretch}{1.2}
\begin{tabular}[t]{C{3.6cm}C{2.0cm}C{2.0cm}}
\toprule
\textbf{Classifier}&\textbf{F1 score without oversampling}&\textbf{F1 score with oversampling}\\
\midrule
\textbf{Support Vector Machine}&88.2\% &\textbf{94.0\%} \\
\textbf{Naive Bayes (Bernoulli Dist.)}&83.8\% &88.2\%\\
\textbf{Naive Bayes (Gaussian Dist.)}&54.2\% &65.6\%\\
\textbf{Logistic Regression}&87.8\% &90.8\% \\
\textbf{Neural Network}&\textbf{89.0\%} &\textbf{94.0\%}\\
\bottomrule
\end{tabular}
\end{table}

\begin{table}[!th]
	\centering
	\caption{Evaluation of the classifiers over the automated account dataset.}
	\label{Table:result_automated}
	\renewcommand{\arraystretch}{1.2}
	\begin{tabular}[t]{C{3.6cm}C{1.2cm}C{0.8cm}C{1.1cm}}
		\toprule
		\textbf{Classifier}&\textbf{Precision}&\textbf{Recall}&\textbf{F1-Score}\\
		\midrule
		\textbf{Support Vector Machine}&\textbf{91\%} &82\% &\textbf{86\%}\\
		\textbf{Naive Bayes (Bernoulli Dist.)}&85\% &68\% &78\%\\
		\textbf{Naive Bayes (Gaussian Dist.)}&51\% &\textbf{98\%} &67\%\\
		\textbf{Logistic Regression}&80\% &70\% &75\%\\
		\textbf{Neural Network}&89\% &84\% &\textbf{86\%}\\
		\bottomrule
	\end{tabular}
	\vspace{10pt}
\end{table}

\section{Conclusions}
In conclusion, detection of the fake and automated accounts which leads to fake engagement in Instagram is studied as a binary classification problem in this paper. To our knowledge, this is the first time for such an analysis over Instagram accounts. Our contributions with this work are: collection of datasets for fake and automated account detection, proposing derived features for fake and automated classification, proposing a cost sensitive feature reduction technique based on genetic algorithms for selecting best  features for the classification of automated accounts, correcting the unevenness in the fake account dataset using the SMOTE-NC algorithm and evaluating several pattern recognition methods over the collected datasets. As a result, SVM and neural network based methods achieved the most promising F1 score for the detection of automated accounts with 86\% and neural network achieved the best F1 score performance with 95\%.  

As a future work, recurrent neural networks can be utilized for the time series user data for a better detection of automated accounts. The biased features in the automated account dataset can be balanced by finding the suitable real users. Fake user detector explained in this paper can also be used for finding the suitable real users in the automated account dataset.


%
%



%
%
%

\bibliographystyle{IEEEtran}
\bibliography{refs}

\end{document}